\documentclass[useAMS,usenatbib]{mn2e}

\usepackage{graphicx}
\usepackage{amsmath}
\usepackage{mathptmx}
\usepackage{times}
\usepackage{epsfig}
\usepackage{color}
\usepackage{mathrsfs}
\usepackage{epstopdf}

\newcommand{\revone}[1]{{ #1}}

\title[Recombination energy in double white dwarf formation]{Recombination energy in double white dwarf formation}
\author[J.L.A. Nandez, N. Ivanova and J.C. Lombardi Jr.]{J.L.A. Nandez$^{1}$\thanks{E-mail:
avendaon@ualberta.ca (JLAN)
}, 
N. Ivanova$^{1}$, and J.C. Lombardi Jr.$^{2}$\\
$^{1}$Department of Physics, University of Alberta, Edmonton, AB, T6G 2E7, Canada\\
$^{2}$Department of Physics, Allegheny College, Meadville, PA 16335, USA}

\newcommand\aj{AJ}%
\newcommand\apj{ApJ}%
\newcommand\apjs{ApJS}%
\newcommand\aap{A\&A}%
\newcommand\aapr{A\&A~Rev.}%
\newcommand\mnras{MNRAS}%
\newcommand\nat{Nature}%
\begin{document}

\date{Draft date: \today}

\pagerange{\pageref{firstpage}--\pageref{lastpage}} 

\maketitle

\label{firstpage}

\begin{abstract}
In  this {\it  Letter} we  investigate the  role of  recombination
energy during  a common  envelope event. We  confirm that  taking this
energy into account  helps to avoid the formation  of the circumbinary
envelope commonly found  in previous studies. For  the first time,
we can model a complete common  envelope event, with a clean compact
double white  dwarf binary system  formed at  the end. The resulting binary orbit is
almost  perfectly circular.  In addition  to considering 
recombination energy,  we also show  that between  1/4 and 1/2  of the
released  orbital energy  is taken  away by  the ejected  material. We
apply this new method to the case of the double-white dwarf
system WD 1101+364,  and we find that the progenitor system  at the start
of the common envelope event consisted of a $\sim1.5M_\odot$ red giant star in a $\sim 30$~day orbit
with a white dwarf companion.
 
\end{abstract}


\section{Introduction}
\label{sec:intro}

The formation of a compact binary system composed of two white
dwarfs (WDs)  is widely  accepted to include  a common  envelope event
(CEE), at least  during the last episode of mass  exchange between the
first-formed WD  and a low-mass red  giant (RG).  Low-mass RGs  have a
well-defined relation  between their core masses  and radii, providing
for DWDs the  best-defined state of a progenitor binary  system at the
onset of the  CEE among all known types of  post-common envelope (CE)
systems.   As  a  result,  DWD  systems  have  served  extensively  as
test-sites for attempts to understand  the physics of CEEs, using both
population synthesis approaches and hydrodynamical methods.

Previous attempts to  model a CEE between  a low-mass RG and  a WD did
not  succeed to  eject  the entire  CE  during three-dimensional  (3D)
hydrodynamical   simulations    \citep[for   most    recent   studies,
  see][]{2012ApJ...744...52P,2012ApJ...746...74R}.  The final state of these
simulations is  that a significant  fraction of the  expanded envelope
remains bound  to the  formed binary,  forming a so-called circumbinary
envelope.   Then almost  no energy  transfer can  take place  from the
binary orbit to that circumbinary envelope. \revone{Observationally,
no circumbinary disk in a post-common envelope system has been found so far.}

It has  been proposed long ago that  recombination  energy of
hydrogen   and   helium   should   play    a   role   during   a   CEE
\citep{1967AJ.....72Q.813L,1967Natur.215..838R,1968AcA....18..255P,1994MNRAS.270..121H,2002MNRAS.336..449H}.
However, until now, this energy was not yet taken into account in 3D
modeling.   While  the  initially  available
recombination energy can be easily comparable to the binding energy of
the remaining bound  envelope \citep[e.g.][]{2012ApJ...744...52P}, the
important question is  {\it when and where} the energy  is released --
to  be useful,  recombination  energy should not  be released too early in the CEE nor in material already ejected,
but  instead in the circumbinary envelope at a time when the recombination energy  is comparable to  the binding energy
of the not-yet ejected material.  In this {\it Letter}, we investigate
if  the inclusion of  recombination energy can help  to remove the circumbinary envelope.   We apply the new approach to
the system WD 1101+364, a  well-measured DWD that has $P_{\rm orb}=0.145$
d and  a mass ratio of  $q=M_1/M_2=0.87\pm0.03$, where $M_1\simeq0.31
M_\odot$ and $M_2\simeq0.36 M_\odot$ are the masses of the younger and
older WDs, respectively \citep{1995MNRAS.275L...1M}.

\section{Initial set up and definitions}
\label{sec:initial}

We anticipate  that the progenitor  of WD  1101+364 was a  low-mass RG
that had a degenerate core of $0.31 M_\odot$. We consider the range of
masses for the  RG donor, $M_{\rm d,1}$, from $1.0 M_\odot$  to $1.8 M_\odot$.
To evolve  the RG  and find the  initial one-dimensional  (1D) stellar
profile,  we use  {\tt TWIN/Star}  stellar code  \citep[recent updates
described  in][]{2008A&A...488.1007G}.  The  stars are  evolved  until
their degenerate He cores have grown close to $0.31 M_\odot$.

\begin{table*}
\begin{minipage}{170mm}
 \caption{Initial conditions}
 \label{tab:init}
 \begin{center}
 \begin{tabular}{lccccccccrcccc}
  \hline
  Model & $M_{\rm d,1}$ & $M_{\rm env,1}$ & $M_{\rm c,1}$ &$R_{\rm rlof}$ & $a_{\rm orb,ini}$ & $P_{\rm orb,ini}$ &$N_{\rm p}$ & $\eta$ & $E_{\rm bind}$&$E_{\rm rec}$&$E_{\rm orb,ini}$&$E_{\rm tot,ini}$&$\lambda$\\
  \hline
  1.0RG0N&0.985 & 0.668 &0.317& 28.21 & 60.11 & 46.57 & 99955 & 0.00&-8.454&2.059&-1.118&-7.513&1.047\\
  1.0RG1N&0.985 & 0.668 &0.317&29.78 & 63.51 & 50.57 &99955 & 1.00&-8.454&2.059 &-1.058&-7.453&0.992\\
  1.0RG2N&0.985 & 0.668 &0.317&31.35 & 66.81 & 54.57 &99955 & 2.00&-8.454&2.059 &-1.006&-7.401&0.942\\
  1.2RG2N&1.195 & 0.877 &0.318&29.48 & 60.74 & 44.00 &99955 & 2.00&-12.328&2.725 &-1.345&-10.945&1.093\\
  1.4RG2N&1.397 & 1.079 &0.319&27.73 & 55.59 & 36.24 &99955 & 2.00&-16.947&3.369 &-1.715&-15.293&1.217\\
  1.5RG2N&1.498 & 1.179 &0.319&25.66 & 50.82 & 30.81 &99955 & 1.65&-20.636&4.038 &-2.015&-18.609&1.267\\
  1.5RG2NP&1.498 & 1.179 &0.318&25.66 & 50.82 & 30.81 &200221& 1.65&-20.345&3.697 &-2.011&-18.659&1.285\\
  1.6RG2N&1.598 & 1.275 &0.323&25.80 & 50.54 & 29.76 &99955 & 2.00&-22.837&4.009 &-2.157&-20.985&1.312\\ 
  1.6RG0S&1.598 & 1.275 &0.323&31.25 & 48.61 & 27.97 &99955 & 0.00&-22.358&3.997 &-2.241&-20.602&1.106\\
  1.7RG2N&1.699 & 1.376 &0.323&22.83 & 44.25 & 23.78 &99955 & 2.00&-28.638&4.338 &-2.619&-26.918&1.356\\
  1.7RG0S&1.699 & 1.376 &0.323&27.58 & 42.97 & 22.67 &99955 & 0.00&-28.003&4.326&-2.694&-26.371&1.148\\
  1.8RG2N&1.799 & 1.481 &0.318&16.34 & 31.37 & 13.86 &99955 & 2.00&-44.167&4.676 &-3.912&-43.404&1.401\\
  \hline
 \end{tabular}
 \end{center}
 \medskip
The models  names are composed  as following: two  digits representing
the RG mass are followed by ``RG'', $\eta$ value is is outermost smoothing length;
then  ``N'' stands  for  non-synchronized and  ``S'' for  synchronized
(``S'') cases.  ``P'' denotes the model with about twice larger number
of particles than in all the  other models. $M_{\rm d,1}$, $M_{\rm env,1}$ and
$M_{\rm c,1}$  are the  total, envelope  and core mass  of the  RG, in
$M_\odot$. $R_{\rm  rlof}$ is the radius  of the donor Roche  lobe, in
$R_\odot$, and $\eta$ describes  the adopted donor's radius definition
(see  \S2).  $a_{\rm orb,ini}$  is  the  initial orbital  separation  in
$R_\odot$,  $P_{\rm   orb,ini}$  is   the  initial  orbital   period  in
days. $N_{\rm p}$ is the total number  of SPH particles that represent the
RG.   $E_{\rm  bind}$,  $E_{\rm  rec}$, $E_{\rm  orb,ini}$  and  $E_{\rm
  tot,ini}$  are  the   binding  energy  of  the   RG  envelope  without
recombination  energy,  the  total  recombination  energy  of  the  RG
envelope,   initial  orbital   energy,  and   initial  total   energy\revone{, defined as the sum of the binding, recombination, and initial orbital energies}, respectively,  in the  units of  $10^{46}$  erg. $\lambda$  is a  star
structure parameter (see Equation \ref{eq:lambda}).
 \end{minipage}
\end{table*}

For 3D  simulations, we use  {\tt StarSmasher} \citep{2010MNRAS.402..105G,2011ApJ...737...49L}, a  Smoothed Particle Hydrodynamics (SPH) code.
Technical details on using this code to treat CE events
can be found  in \cite{0004-637X-786-1-39}.  A 1D  stellar profile is
imported to {\tt StarSmasher},  where an initial stellar model represented  by a
certain number of particles $N_{\rm p}$ is generated via a {\it relaxation} process.  The
core of a RG  is modeled as a point mass --  a “special particle” in SPH
that  interacts only gravitationally with other  particles.  Because the centre of the giant is not fully resolved,   the   core  mass,   $M_{\rm  c,1}$,   is
slightly    more   than   in  the  1D    code   (see
Table \ref{tab:init}  for this  and other initial  values).  This
ensures a proper matching of stellar  profiles of 3D envelopes with 1D
stellar  profiles.   The  envelope  mass  in  a  3D  star  is  $M_{\rm
env,1}=M_{\rm d,1}-M_{\rm c,1}$.

In a  3D star, the  radius of the  star, $R_{\rm SPH}$,  can not be  defined as
uniquely as the  photospheric radius of the 1D star  \citep[for a thorough
  discussion,  see][]{{0004-637X-786-1-39}}.   The stellar radius can  be
parameterized as  $R_{\rm SPH}=R_{\rm out}+\eta h_{\rm  out}$, where $R_{\rm
  out}$ is the  position of the outermost particle  and $h_{\rm out}$
is the smoothing length of that  particle.  The parameter $\eta$ can range between 0
(in  this case,  some mass  will be  found above  $R_{\rm
  SPH}$) to 2 (with all  mass contained  within  $R_{\rm  SPH}$).   In
addition, we note that a  synchronized  giant  is expected  to  attain   a larger radius after
relaxation than a non-synchronized giant.

The initial orbital separation,  $a_{\rm orb,ini}$, for the non-synchronized cases, is found from
the assumption  that $R_{\rm  SPH}$ is  equal to  the Roche  lobe (RL)
overflow  radius,   $R_{\rm  rlof}$,   and  using   the  approximation
by \citet{1983ApJ...268..368E}.   The initial orbital  period, $P_{\rm
orb,ini}$ is found assuming a Keplerian orbit. 
For the  synchronized cases, the orbital period and separation  are found at the moment when the RG overflows its Roche lobe 
\citep[see \S2.3 of][]{2011ApJ...737...49L}.

\textbf{Equations of state (EOS). }
The standard  EOS (SEOS) in {\tt StarSmasher} is analytical and  includes radiation pressure
and ideal gas contributions.  To  take into account recombination energy,  we need another
prescription for the  EOS.  Because we evolve the  specific internal energy
$u_i$  and   density  $\rho_i$  for  each   particle  (among  other
variables), we prefer an EOS  that uses  $u_i$ and  $\rho_i$ as
independent variables.   However, such an analytical expression  does not
exist in simple form when  we consider recombination/ionization of  atoms. Therefore,
we  are bound  to use  a  tabulated EOS  (TEOS) which  uses $u_i$  and
$\rho_i$ to provide the  gas pressure $P_{{\rm gas},i}$, temperature
$T_i$, specific entropy $s_i$, etc.

We   use   the    MESA-EOS   module   to  calculate   such
tables \citep[see \S4.2 of][]{2011ApJS..192....3P}.   The core of a RG
is modeled  as a  point mass,  and the  rest of  the star  has uniform
composition.  Hence, only  one table with a single  set of composition
for H, He, and metals needs to  be generated for each  RG.  The tables
that we generate  operate in $9.84\leq \log u  [{\rm erg\ g^{-1}}]\leq
19.0$  and $-14\leq  \log \rho  [{\rm g\  cm^{-3}}]\leq 3.8$.   When a
particle  has a  density or  specific internal  energy outside the
limits of  our tables,  we switch  to the SEOS.

\textbf{Energy formalism. }
The energy formalism compares the donor's envelope binding energy $E_{\rm bind}$
with the orbital energy before, $E_{\rm orb,ini}$, and after the CEE, $E_{\rm orb,fin}$
\citep{1984ApJ...277..355W,1988ApJ...329..764L}:
\begin{equation}
E_{\rm bind}=\alpha_{\rm bind} (E_{\rm orb,fin}-E_{\rm orb,ini}) \equiv \alpha_{\rm bind} \Delta E_{\rm orb} \ . 
\label{eq:standardCE}
\end{equation}
Here  $\alpha_{\rm bind}$ is the fraction of the released orbital energy used to expel the CE, $0\le\alpha_{\rm bind}\le 1$. 
The binding energy of the donor's envelope, in its standard definition, is
\begin{equation}
 E_{\rm                 bind}                 =                 \sum_i
 m_i\left(\phi_i+\frac{3}{2}\frac{kT_i}{\mu_im_{\rm
 H}}+\frac{aT_i^4}{\rho_i}\right),
\end{equation}

\noindent where  $m_i$, $T_i$,  $\rho_i$,  $\phi_i$ and  $\mu_i$  are the  mass,
temperature, density, specific gravitational potential energy, and mean
molecular mass,  respectively, for each  particle $i$. The constants $k$, $a$,  and $m_{\rm H}$
are the Boltzmann constant, radiation constant, and hydrogen atom mass,
respectively, while $\phi_i$        is         calculated        as
in  \citet{1989ApJS...70..419H}.   Note  that $E_{\rm  bind}$  in  its
standard definition does not include recombination energy.

The binding energy of the donor's envelope is frequently parameterized  
using a parameter $\lambda$, defined as  \citep{1990ApJ...358..189D}
\begin{equation}
 \lambda \equiv -\frac{GM_{\rm d,1}M_{\rm env,1}}{E_{\rm bind} R_{\rm rlof}}.
\label{eq:lambda}
\end{equation}
Here $G$ is the gravitational constant.  For low-mass giants, 
$\lambda$ is known to have a value close to one, 
and we obtain similar results.

We find the orbital energy  of the binary system according to
\begin{equation}
\label{eq:orbenergy}
 E_{\rm orb}= \frac{1}{2}\mu |V_{12}|^2
   +\sum_i   \frac{1}{2}m_i\phi_i-\sum_j   \frac{1}{2}m_j\phi_j^{\rm
   RL_1}-\sum_k \frac{1}{2}m_k\phi_k^{\rm RL_2},
\end{equation}

\noindent where $\mu=M_1M_2/(M_1+M_2)$ is the reduced mass, and $\vec{V}_{12}=\vec{V}_1-\vec{V}_2$ is the relative velocity of the two stars. The first term gives the orbital kinetic energy.
The second term  is the total gravitational energy  of the
binary, with the sum being over  all particles $i$ in  the binary. The  third and
fourth terms correspond to the removal of the self-gravitational energy of the donor (the sum
being over particles $j$  in star 1) and of the WD (the sum being over particles $k$
in star 2, initially only the WD), respectively: the remaining gravitational energy is then just the orbital contribution.

\textbf{Recombination energy. }
\revone{In our treatment, the internal energy provided by the TEOS includes contributions from ideal gas, radiation, and the recombination energy for H, He, C, O, N, Ne, and Mg \citep[see \S4.2 of][]{2011ApJS..192....3P}. 
The recombination energy 
can be extracted}  as
\begin{equation}
 E_{\rm rec}  = \sum_i m_i\left(u_i-\frac{3}{2}\frac{kT_i}{\mu_im_{\rm
 H}}-\frac{aT_i^4}{\rho_i}\right)\equiv \alpha_{\rm  rec}\Delta E_{\rm
 orb}, \label{eq:recene}
\end{equation}

\noindent  where $u_i$ is the SPH specific internal energy of particle $i$. 
Values of $E_{\rm rec}$, as expected, scale well with the mass of the envelope. 
Note  that here  we  introduce important  new parameter,  $\alpha_{\rm
rec}$ -- the ratio between the recombination energy and the released orbital
energy.

\textbf{Total energy. } The initial total available  energy, 
$E_{\rm tot,ini}$, is 
\begin{equation}
 E_{\rm tot,ini}=E_{\rm orb,ini}+E_{\rm bind}+E_{\rm rec}.
 \label{eq:etotin}
\end{equation}
This quantity is conserved during the evolution of all our models.

\textbf{Bound and unbound material. } For each particle, its total energy is defined as  $E_{{\rm tot,}i}\equiv
0.5 m_iv_i^2 + m_i\phi_i+m_iu_i$, where the first, second and third terms are the kinetic, potential, and internal energies, respectively.  If  the
particle has negative total energy, it is bound to the binary.
In this case, if  the particle is located  outside of either RL,  the  particle  is in  the  circumbinary
region.   Accordingly,  we   classify   the  particles   to   be  in   (i)
the  \textit{ejecta},  $m_{\rm  unb}$ --  the  particles  with
positive energy, (ii) the  \textit{circumbinary}, $m_{\rm cir}$ -- 
the matter bound to the binary  but outside of the two RLs, and
(iii) \textit{binary},  $m_{\rm bin}$ -- the particles are
inside either of the two RLs.

The total energy of the unbound material at infinity is computed when 
the unbound mass is in a steady state after the CEE:
\begin{equation}
 E_{\rm tot,unb}^\infty =  \sum_i E_{\rm tot,i}^{\rm unb}=-\alpha_{\rm unb}^\infty\Delta E_{\rm orb}\ .  \label{eq:totunb}
\end{equation}
\noindent Here we introduce $\alpha_{\rm unb}^{\infty}$ -- the energy
taken  away by the  unbound material in units of  the released
orbital  energy.  Note  that  in the  standard  energy formalism  this
quantity is always assumed to be zero.

\textbf{Angular momentum budget. }
 We calculate the orbital angular momentum 
\begin{equation}
 \vec{J}_{\rm orb} \equiv \mu \vec{R}_{12}\times \vec{V}_{12},
\end{equation}
where $\vec{R}_{12}=\vec{R}_1-\vec{R}_2$ is the displacement from star 2 to star 1. We note that the magnitude $J_{\rm orb}=|J_{\rm orb,z}|$, where the $z$-direction is perpendicular to the orbital plane.
An outcome of a CEE can be characterized by how much orbital angular momentum is lost. 
We provide the $\gamma$-parameter  
\citep{2000A&A...360.1011N,2005MNRAS.356..753N} as a way of quantifying angular momentum loss in our simulations: 
\begin{equation}
 \gamma = \frac{M_1+M_2}{M_{\rm unb}}\frac{J_{\rm orb,ini} -J_{\rm orb,fin}}{J_{\rm orb,ini}}.
 \label{eq:gamma}
\end{equation}

\section{Formation of a DWD through a  CEE}
\label{sec:dwdce}

\begin{figure}
 \includegraphics[scale=.4]{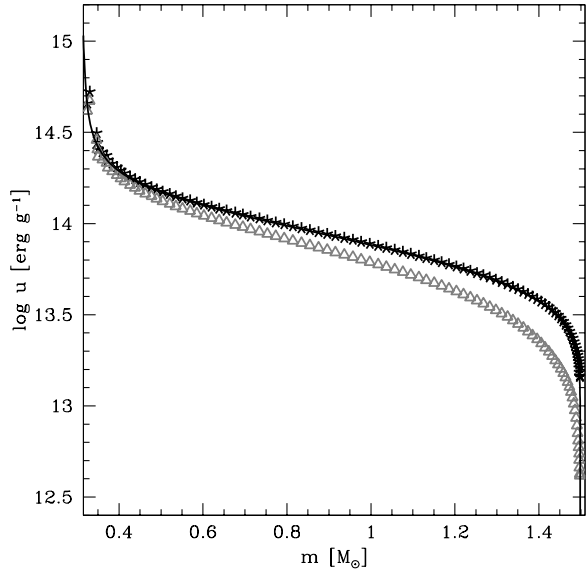}
 \caption{Specific internal energy $u$ profiles for the model 1.5RG2N. 
The black asterisks and gray triangles correspond to relaxed $u$ profiles 
for TEOS and SEOS, respectively. The black solid line corresponds to 
the $u$ profile from the stellar code.}
 \label{fig:fig2}
\end{figure}

{\bf Comparison between the two EOSs.} We compare the results of simulations with our two different EOSs
 using the model 1.5RG2N.
In both cases, the initial  relaxed stars
have  SPH  profiles that match  very well  the 1D
stellar profiles for
pressure,  density,  and  gravitational  potential.  However, this  is not  the case  for the  specific internal
energy $u$  (see Figure \ref{fig:fig2}): clearly only
the TEOS model matches the desired 1D stellar  profile.  As expected, the mismatch  between the relaxed
profile with  SEOS and  the stellar  one is indeed due to neglecting recombination
energy.

We find that the SEOS fails to unbind  the envelope in our CE simulations. Only about
50\% of the envelope becomes unbound: the circumbinary matter does
not interact  with the formed binary  at all, making it  impossible to
eject  the  entire  envelope.  This result  is  consistent  with  the findings of
previous studies \citep{2012ApJ...744...52P,2012ApJ...746...74R}. On the other hand,
the TEOS simulation clearly makes use of the
recombination energy and ejects the envelope entirely.
For all other models presented in the {\it Letter}, we use the TEOS.

{\bf  Masses.}  At the  end  of  the  simulations,  we form  a  binary
consisting of  $M_1$ and  $M_2$ (see Table  \ref{tab:DWD} for  all the
outcomes).  The unbound  material $M_{\rm unb}$ is at  least 99.8\% of
the initial  envelope. A few, usually  less than 10, SPH  gas particles
remain bound  to the  newly formed  binary, been  bound to  either the
newly  formed WD,  or the  old  WD.  This  explains why  $M_1$ can  be
slightly larger than $M_{\rm c,1}$,  and similarly why $M_2$ can
exceed  slightly $0.36  M_\odot$.  There  is no  circumbinary envelope
left in  all simulations with the  TEOS.  In all our  simulations, the
final mass  ratio $q$  ranges between  $0.88-0.90$, consistent
with the observational error for WD 1101+364.

\begin{table*}
\begin{minipage}{165mm}
 \caption{Energies and masses}
 \label{tab:DWD}
 \begin{center}
 \begin{tabular}{lcccccccccccc}
  \hline
  Model & $M_{\rm unb}$ & $M_{1}$ & $M_2$ & $E_{\rm kin,unb}^{\infty}$ & $E_{\rm int,unb}^{\infty}$ & $E_{\rm pot,unb}^{\infty}$&$E_{\rm tot,unb}^{\infty}$&$E_{\rm orb,fin}$ & $E_{\rm bound}$ & $E_{\rm tot,fin}$&$\Delta E_{\rm orb}$ \\
  \hline
  1.0RG0N&0.663&0.322&0.360&3.645&0.473&-0.025&4.093&-10.992&-0.615&-7.514&-9.874\\
  1.0RG1N&0.663&0.322&0.360&4.123&0.295&-0.015&4.403&-11.278&-0.582&-7.457&-10.220\\
  1.0RG2N&0.663&0.322&0.360&4.081&0.543&-0.024&4.600&-11.469&-0.531&-7.400&-10.490\\
  1.2RG2N&0.872&0.323&0.360&4.604&0.629&-0.041&5.192&-15.504&-0.639&-10.951&-14.159\\
  1.4RG2N&1.079&0.319&0.360&6.790&0.907&-0.094&7.603&-22.911&-0.005&-15.313&-21.196\\
  1.5RG2N&1.178&0.319&0.361&6.089&0.917&-0.096&6.910&-25.484&-0.026&-18.600&-23.469\\
  1.5RG2NP&1.178&0.320&0.360&7.415&1.407&-0.159&8.663&-26.969&-0.366&-18.665&-24.958\\
  1.6RG2N&1.274&0.323&0.362&5.623&1.812&-0.440&6.995&-27.741&-0.244&-20.990&-25.584\\
  1.6RG0S&1.274&0.323&0.362&5.603&1.692&-0.381&6.914&-27.228&-0.309&-20.623&-24.987\\
  1.7RG2N&1.370&0.323&0.366&5.854&2.042&-0.417&7.479&-33.692&-0.715&-26.928&-31.073\\
  1.7RG0S&1.373&0.323&0.363&5.032&2.061&-0.610&6.483&-32.417&-0.466&-26.400&-29.723\\
  1.8RG2N&1.478&0.318&0.362&8.333&1.675&-0.371&9.637&-52.873&-0.171&-43.407&-48.961\\
  \hline
 \end{tabular}
 \end{center}
 \medskip
 $M_{\rm unb}$, $M_1$,  and $M_{2}$ are the unbound,  stripped RG core
 and  old   WD,  in  $M_\odot$.  \revone{$E_{\rm   kin,unb}^{\infty}= \sum_i m_i^{\rm unb} v_i^2/2$, 
 $E_{\rm   int,unb}^{\infty}=\sum_i m_i^{\rm unb} u_i$,   $E_{\rm   pot,unb}^{\infty}=\sum_i m_i^{\rm unb} \phi_i$},   and   $E_{\rm
   tot,unb}^{\infty}$  are  kinetic,  internal,  potential  and  total
 energies,  respectively,  for  the   unbound  material.  $E_{\rm orb,fin}$  is  the  orbital energy  after  the CEE.
 \revone{$E_{\rm bound}$ is the total energy  of the particles that remained bound to the binary. 
$E_{\rm  tot,fin}$ is  the total  energy of all the particles. All energies are in $10^{46}$ erg.}
 \end{minipage}
\end{table*}
{\bf Energies.} The total  energy at the end  of the simulation is  distributed in the
``binding''  energy of  the  gas bound  to the  binary,
$E_{\rm  bound}$, the  final orbital  energy of  the binary,  $E_{\rm
orb,fin}$, and the  total energy of the unbound  material at infinity,
$E_{\rm tot,unb}^{\infty}$:
\begin{equation}
 E_{\rm tot,fin}=E_{\rm orb,fin}+E_{\rm bound}+E_{\rm tot,unb}^{\infty}.
\label{eq:totefinal}
\end{equation}

\noindent 
We have  compared the initial and  the final total energies  and found
that the error is less than 0.11\% in all our simulations.

$E_{\rm tot,unb}^{\infty}$ is  composed of $E_{\rm kin,unb}^{\infty}$,
$E_{\rm int,unb}^{\infty}$, $E_{\rm pot,unb}^{\infty}$ -- the kinetic,
internal and potential energies of the unbound material, respectively.
We note that $E_{\rm kin,unb}^{\infty}$  is the dominant energy in the
unbound material, though  the internal energy of  the unbound material
at  the end  of  the  simulations is  also  non-negligible (see  Table
\ref{tab:DWD}).

We present $E_{\rm  bound}$ for completeness, but the fate --  accretion or
ejection --  of the
several  particles that  remain bound  to the  binary can not be resolved by the numerical method we use; on the
timescale of  our simulation they  stay in an  orbit within the  RL of
their  stars.   This  energy  includes  the  kinetic, internal,  potential  and
recombination energies for these several SPH gas.

We should clarify  that $E_{\rm orb,fin}$ does not have  to match with
the  two-body  approximation,  namely  $E_{\rm  orb}=-GM_1M_2/(2  a_{\rm
  orb})$.   In  the   latter,  the  potential  assumes   a  form 
$\phi\propto1/r$,  while  our  code includes  the softened form  as
described in the appendix of \cite{1989ApJS...70..419H}. When the  separation between
the two  SPH special particles is  more than two smoothing  lengths, the
potential reduces to the Keplerian form. However, this separation is less
than two smoothing  lengths for the point particles  after the CEE,
and the  potential is softened accordingly.
The
difference in orbital energy between the two  methods varies from about 3\% (for 1.0RG0N) to 14\% (for 1.8RG2N), with the Keplerian values being closer to zero.  The initial orbital energy,  given by
Equation  \ref{eq:orbenergy}, is  the  same as in the
two-body approximation.  

\begin{table*}
\begin{minipage}{175mm}
 \caption{Orbital parameters}
 \label{tab:DWDj}
 \begin{center}
 \begin{tabular}{lcccccccccccc}
  \hline
  Model & $J_{\rm orb,ini}$ & $J_{\rm orb,fin}$ & $\gamma$& $r_{\rm p}$ & $r_{\rm a}$ & $a_{\rm orb,fin}$ & $P_{\rm orb,fin}$ & $e$ &$\alpha_{\rm bind}$&$\alpha_{\rm rec}$&$\alpha_{\rm unb}^{\infty}$\\
  \hline
  1.0RG0N& 14.340 & 1.188  & 1.861&2.015&2.115&2.065&0.416&0.024&0.855&-0.208&0.431 \\
  1.0RG1N& 14.741 & 1.168  & 1.868&1.965&2.074&2.020&0.403&0.027&0.827&-0.201&0.431 \\
  1.0RG2N& 15.119 & 1.157  & 1.873&1.947&2.036&2.000&0.397&0.022&0.808&-0.197&0.440 \\
  1.2RG2N& 16.262 & 0.987  & 1.670&1.520&1.532&1.526&0.264&0.004&0.871&-0.192&0.367 \\
  1.4RG2N& 17.116 & 0.759  & 1.557&1.070&1.089&1.080&0.158&0.009&0.800&-0.159&0.359\\
  1.5RG2N& 17.062 & 0.709  & 1.512&0.953&1.003&0.978&0.134&0.026&0.879&-0.172&0.294\\
  1.5RG2NP&17.062 & 0.719  & 1.511&0.891&0.924&0.908&0.122&0.018&0.815&-0.148&0.347\\
  1.6RG2N& 17.685 & 0.678  & 1.479&0.880&0.948&0.914&0.122&0.037&0.893&-0.157&0.273\\
  1.6RG0S& 17.392 & 0.690  & 1.477&0.912&0.947&0.930&0.126&0.019&0.895&-0.160&0.277\\
  1.7RG2N& 17.151 & 0.610  & 1.449&0.746&0.771&0.758&0.092&0.016&0.922&-0.140&0.241\\
  1.7RG0S& 16.953 & 0.624  & 1.444&0.776&0.791&0.784&0.097&0.009&0.942&-0.146&0.218\\ 
  1.8RG2N& 14.932 & 0.446  & 1.417&0.464&0.493&0.479&0.047&0.030&0.902&-0.096&0.197\\
  \hline
 \end{tabular}
 \end{center}
 \medskip
The orbital angular momentum $J_{\rm  orb,ini}$ and $J_{\rm  orb,fin}$ for the initial and final binary,
 respectively, in units of $10^{52}\ \rm
 g\ cm^2\ s^{-1}$. The parameter $\gamma$ is defined in
 Eq.~\ref{eq:gamma}.  The closest and farthest orbital separations are $r_{\rm p}$ and $r_{\rm a}$, respectively, while $a_{\rm  orb,fin}$ is  the
semimajor axis (all in $R_{\odot}$). The orbital period $P_{\rm orb,fin}$ is given  in  days, and  $e$  is  the  eccentricity  of  the
 orbit.  The energy fractions $\alpha_{\rm  bind}$,  $\alpha_{\rm rec}$,  and  $\alpha_{\rm
   unb}^\infty$   are defined   in
 Eq.      \ref{eq:standardCE},      Eq.      \ref{eq:recene},      and
 Eq. \ref{eq:totunb}, respectively.
 \end{minipage}
\end{table*}

Because energy is well conserved, we can equate Equations \ref{eq:etotin}
and   \ref{eq:totefinal}.    For   that,   we   also   use   Equations
\ref{eq:standardCE}, \ref{eq:recene}, and \ref{eq:totunb}.  \revone{If we neglect $E_{\rm bound}$}, we can re-write the conservation of energy
in fractions of the change in the orbital energy:
\begin{equation}
 \alpha_{\rm bind}+\alpha_{\rm rec}+\alpha_{\rm unb}^\infty \approx1.
\end{equation}
We find  that this is indeed  the case in our simulations (see Table~3), 
\revone{and that the deviation from 1  is due to $E_{\rm bound}$: the maximum deviation occurs in 1.0RG0N ($\sim7.8\%) $ and the minimum in 1.4RG2N ($\sim0.03\%$).}   Note that if
$\alpha_{\rm  rec}=\alpha_{\rm unb}^\infty=0$,  the previous  equation
reduces to  the standard  energy formalism.   However, values  of both
$\alpha_{\rm  rec}$ and  $\alpha_{\rm unb}^\infty$  are non-negligible
and comparable  to $\alpha_{\rm bind}$.  We  emphasize that previously
it had been  anticipated only that  $\alpha_{\rm bind}$  is somewhat
less than 1,  and we provide new  improved constraints.  Unfortunately,
this is not yet a final solution of the problem as $\alpha_{\rm unb}^\infty$ can
not be  easily predicted for  any system---a subject  of our
future studies.

{\bf Orbital  angular momenta.  }  We find  that the  ejected material
takes away more than 90\% of the initial angular  momentum of the
binary.  Values  of $\gamma$ vary  between 1.42 and 1.87.   This large
range of values unfortunately does not allow the obtained values
of $\gamma$ to be useful for predicting the final parameters in a population
synthesis   for  all   possible   DWD   systems  \citep[for   details,
see][]{2013A&ARv..21...59I}.

\textbf{Final orbital parameters. }
We  find  the final  orbital  separation  as $a_{\rm  orb,fin}=(r_{\rm
a}+r_{\rm p})/2$, where $r_{\rm p}$ is the periastron, and $r_{\rm a}$
is the apastron.  We ensure that these two quantities, $r_{\rm p}$ and
$r_{\rm a}$,  do not change  with time at  the moment when  we extract
them from  the simulations.  We calculate  the final orbital period $P_{\rm  orb,fin}$  of the
binary from Kepler's third law and the
eccentricity of the post-CE orbit as $e=(r_{\rm a}-r_{\rm p})/(2a_{\rm
orb,fin})$.   The latter  is small  in  all the  models, showing  that
post-CE orbits  are almost circular \citep[in  previous studies, where
the  ejection of  the CE  was incomplete,  the final  eccentricity was
larger, 0.08 or more, e.g.][]{2012ApJ...746...74R}.

\begin{figure}
 \includegraphics[scale=.4]{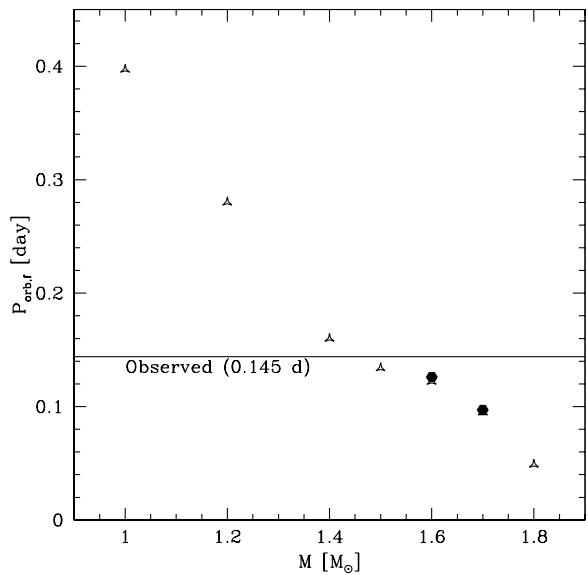}   
\caption{Final  orbital
 periods versus initial mass of a RG. Triangles represent non-synchronized RGs,
and circles represent synchronized RGs. \revone{Note: to compare alike cases, we show only the $\eta=2$ cases with our standard resolution. 
Different $\eta$'s give similar outcomes (see Table~3). }}
 \label{fig:fig4}
\end{figure}

Figure \ref{fig:fig4}  shows the final orbital  periods plotted versus
initial RG.  We see that,  as expected, the more  massive the
star is, the tighter the orbit gets.  We find also that the final orbital
period  for  the  non-synchronized  and synchronized  cases  are  very
similar (for the final state of the binary system, the only change due
to  syncronization was  observed  for the  final eccentricity,  albeit
final eccentricity  is small in all  the cases).  We conclude that our
best progenitor for WD 1101+364 is  a $1.4-1.5M_\odot$ RG.

\section{Conclusions}

To  understand  the energy  budget  during  a  CEE  leading to  a  DWD
formation,  we   perform  non-  and  synchronized   3D  hydrodynamic
simulations with two EOSs.  We confirm that taking into account
recombination energy leads to a full ejection of the RG's envelope and
the formation of a non-eccentric binary system,  whilst if we do not take
recombination energy into account, we obtain result similar  to previous studies and only 
half of the RG's envelope is  ejected.  The most important consideration appears not to
be the value of the  available recombination energy, but where and
when  this energy  is  released.
\revone{Indeed, ionized material forms the circumbinary envelope initially.
Recombination then takes places there, while the circumbinary envelope continues to expand.
This results in the ejection of the circumbinary envelope and effectively of all the common envelope material.}
If instead the recombination  energy had been  released too  early, the  simulations
would  have  ended up  with  unexpelled  circumbinary envelope  as  in
previous studies.   In addition, we  find that considering  a complete
synchronization   versus  non-synchronized   case   does  not   change
noticeably the final results.

 We    introduce   a    modification   of    the   standard    energy
 formalism  \citep{1984ApJ...277..355W,1988ApJ...329..764L}, with  the
 parameters describing  the use  of the  recombination energy  and the
 unbound material energy.  The first  one can be found from initial
 stellar models, but the latter requires  3D simulations. For our  set of
 models, $\alpha_{\rm unb}^\infty$ has values  from about 0.2 to about
 0.44.  However  to generalize  the  result  and  make it  useful  for
 population synthesis  one needs to  make a thorough  parameter study;
 this is the subject of our future studies.

As expected,  we find that the  more massive the parent RG star  is, the tighter
the final orbit gets.   We do  not find  that the  initial synchronization
affects  the final  period but instead  only changes  the energy  and angular
momentum carried  away by  the ejecta, presumably
shaping the post-CE nebula.  We also  find that our binaries end up
with  an eccentricity  smaller than  0.04---a result that  has been  expected
theoretically but  not yet produced in simulations.

We  applied our  method  to  the case  of  WD  1101+364, a  well-known
DWD   \citep[see][]{1995MNRAS.275L...1M}.    We  inferred   that   its
progenitor binary could  have been composed of  a $1.4-1.5M_\odot$ RG
and a $0.36M_\odot$ WD companion, with $P_{\rm orb,ini}\approx31-33$ days.

\section*{Acknowledgments}

JLAN acknowledges CONACyT for its support. 
NI thanks NSERC Discovery and Canada Research Chairs Program. 
JCL is supported by National Science Foundation (NSF) grant number AST-1313091.
This research has been enabled by the use of computing resources provided 
by WestGrid and Compute/Calcul Canada.


\label{lastpage}

\end{document}